\documentclass[11pt]{article}
\usepackage{blois2002,epsfig}
\def\Journal#1#2#3#4{{#1} {\bf #2}, #3 (#4)}

\def\NCA{\em Nuovo Cimento}

\def\NPB{{\em Nucl. Phys.} B}
\def\PLB{{\em Phys. Lett.}  B}
\def\PRL{\em Phys. Rev. Lett.}
\def\PRD{{\em Phys. Rev.} D}

\def\be{\begin{equation}}
\def\ee{\end{equation}}
\def\bea{\begin{eqnarray}}
\def\eea{\end{eqnarray}}

\begin{document}
\hfill{\sf CERN-TH/2002-218}
\vspace*{4cm}
\title{VAST ANTIMATTER REGIONS AND SCALAR CONDENSATE BARYOGENESIS}

\author{ D. KIRILOVA~\footnote{Regular Associate of Abdus Salam ICTP, 
Trieste}, M. PANAYOTOVA, T.VALCHANOV }

\address{Institute of Astronomy, BAS, Sofia and Theory Division, 
CERN, Geneva;\\ CA, Sofia; 
  Institute of Astronomy, BAS, Sofia, Bulgaria}

\maketitle\abstracts{
The possibility of natural and abundant creation of antimatter
in the Universe in a SUSY-baryogenesis model with a scalar field
condensate is described. This scenario predicts  vast quantities of
antimatter, corresponding to galaxy and galaxy cluster  
scales
today,  separated from the matter ones by baryonically empty voids. 
Theoretical and observational constraints on such antimatter regions 
are discussed.
}

\section{Antimatter in the Universe  -- Observational Status}

Is our Universe globally baryonic or the
observed
baryon asymmetry is just a local characteristic? We do not know the
answer, yet.
The observed value of the  baryon asymmetry in our local vicinity is:
\begin{center}
$\beta=(N_B-N_{\bar{B}})/ N_\gamma\sim 10^{-9}-10^{-10}$, 
\end{center}
where  $N_B$ and  $N_{\bar{B}}$ are the baryon and antibaryon number
densities and $N_\gamma$ is
the photon density. 

The  available  cosmic ray (CR) and  gamma ray data  points 
to a
{\it strong predominance of matter over antimatter in our Galaxy:}
\ \\
{\bf Experimental search for antinuclei and $\bar{p}$ in CR} were
conducted on high-altitude balloons and on spacecraft.
 $\bar{p}$ detected  in primary
cosmic radiation over energies
$0.1-19$ GeV are with  negligible numbers, their  ratio to
protons consists  few $10^{-5}$ for
energies lower than 2 GeV and a few $10^{-4}$ for higher energies.
They can be totally due to interactions of  primary CR particles with
the interstellar medium.

No antinuclei were observed. 
The upper limit on the ratio of
antihelium-to helium flux from BESS flights~\cite{saeki98} is
$1.7\times10^{-6}$;
at energies $0.1-8.6$
GeV/nucl. obtained in
balloon experiments~\cite{ormes97} is $8\times10^{-6}$;
 from BESS magnetic rigidity spectrometer in rigidity region 1 to
16 GV $3.1\times10^{-6}$, i.e. the model-independent upper limit on the
antihelium flux, is $6\times10^{-4}$ m$^{-2}$sr$^{-1}$s$^{-1}$,
and for nuclei with $Z>3$ within energy range $1-15$ Gev/nucl is
$8\times10^{-5}$.
The upper limit  from Alpha Magnetic
Spectrometer is $1.1\times10^{-6}$ ($95\%$ C.L.) in the rigidity range
$1-140$
GeV (assumed that the $\bar{He}$ spectrum is with the same shape as
the $He$ one)~\cite{steuer01,alpat00}.
The search will continue in future
AMS and  PAMELA missions
as far as an antinucleus detection would be  a certain signature for
antimatter BBN
or antistars, because the secondary flux for antinuclei is expected to be
extremely low~\cite{orloff}.

Thus, {\it CR  results
indicate that there is no antimatter objects  within a radius 1 Mpc.}

However, {\it the data  are not  definite for larger scales.} 
In Fig.1 we present all the published BESS data~\cite{BESS}, namely the 
antiproton 
spectrum for 1995, 1997-2000. As far as the measurements of $\bar{p}$ 
spectrum at energies above a few GeV are free of uncertainties due to 
secondary $\bar{p}$ production  and solar modulation effects, we present 
also  CAPRICE~\cite{CAPRICE} and 
MASS~\cite{Maeno00} $\bar{p}$ high energy measurements. 
The curves present the theoretical predictions for the secondary 
$\bar{p}$ by Bieber et al.~\cite{Bieber}, 
and Bergstrom (from ref.~\cite{CAPRICE}),  
 calculated within contemporary two-zone diffusion models for the 
corresponding level of solar activity.
The uncertainties due to propagation range between $10\%$ and $20\%$ 
depending on the part of the spectrum~\cite{donato}. 

\begin{figure}
\hbox{\vspace{-0.1cm}}
\mbox{\hspace{-2cm}}\psfig{figure=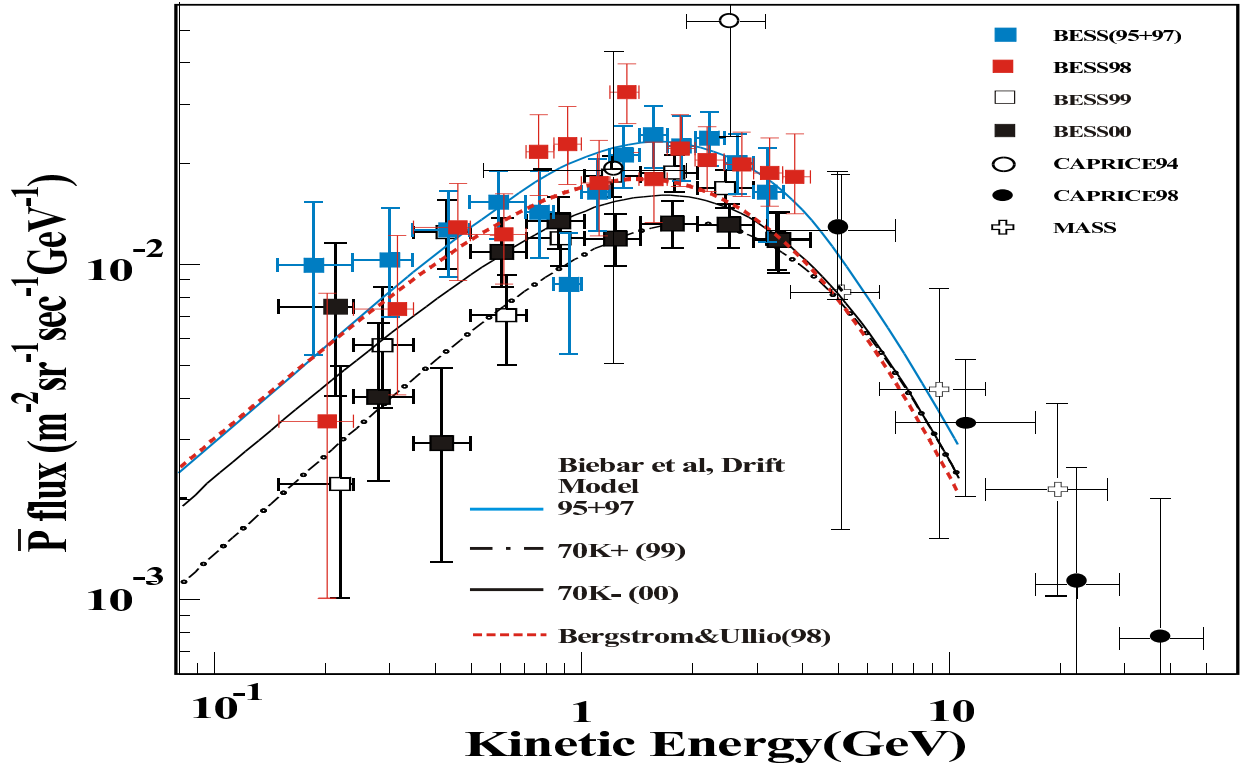,height=15cm,width=18cm}

\hbox{\vspace{-5cm}}
\caption{BESS 1995-2000 antiproton spectrum at the 
top of the atmosphere 
and CAPRICE and MASS data. The curves represent the theoretical 
calculations for secondary $\bar{p}$ for the corresponding solar activity 
level.}
\hbox{\vspace{-1cm}}
\end{figure}

Although the measured   $\bar{p}$-flux and
its spectrum is in agreement with the predicted ones for secondary
particles, the data do not exclude a primary component.
There are even some hints for $\bar{p}$ excess:

(a) An interesting study of the antiproton spectrum through years with
solar minimum (1995, 1997) and maximum (1998), showed that for the low
energy region of the spectrum the agreement during solar minimum is less
consistent than for the maximum.
As far as  $\bar{p}$ from primary sources are suppressed as solar
activity increases, while secondary $\bar{p}$ spectrum is affected
modestly, the results were interpreted in favor of a primary
$\bar{p}$~\cite{Maeno00}.

(b) Slightly excessive $\bar{p}$ fluxes, relative to the theoretical
calculations were found  during solar minimum 
in the analysis by Orito and by Matsunaga~\cite{BESS} 
 for the energies below 0.5 GeV.

It is interesting to provide similar analysis including the available new 
BESS data for 1999 and 2000 years of maximum of solar activity and try to 
limit the possible extragalactic component of $\bar{p}$ and, hence, the 
primary $\bar{p}$.
Our  preliminary analysis of all the available BESS data  
 do not find such trend, however~\cite{kvp02}.
In case an excess  of low energy antiprotons will be disfavored by 
 the  data, still high energy spectrum may be studied,  
having in mind also that this part of the spectrum is  free of 
uncertainties 
due to solar modulation effects, and 
 in this energy range all calculations of secondary
$\bar{p}$ are consistent with each other.

(c) Two antiproton events with the highest energy antiproton were measured 
at a kinetic energy 43 GeV, between 29 and 49 GeV, compared with an
expected number from secondaries only 0.2 to 0.4 events~\cite{boezio02}.
 
So, a fraction
of the observed $\bar{p}$ may well be CR from distant antigalaxies.

In conclusion, {\it the statistical sample of  $\bar{p}$ presently 
available is very limited,
so that a primary component cannot be ruled out with high significance,
even in case the propagation parameters were known.}
Besides,  CR at the rigidities accessible to
current
antimatter experiments should be strongly suppressed by galactic, cluster
and intergalactic magnetic fields~\cite{ormes97}.

 {\bf Gamma rays data, interpreted as a result from annihilation} 
provides  observational constraints  on the antimatter fraction of 
different structures~\cite{stecker,steigman76,wolfendale}. 
No evidence for annihilation features due to contacting matter and
antimatter in the period $z<100$
was found in the cosmic gamma ray background.
The measurements of the gamma ray flux in the MeV region exclude 
significant amounts of antimatter up to the distance of 
 galaxy cluster scales $\sim 10-20$ Mpc~\cite{stecker71}. 
Hence, it is interesting to explore baryogenesis models predicting large 
antimatter structures. 

The  analysis of the relic gamma rays contribution  
from early annihilation to the cosmic diffuse gamma spectrum  gave 
the limit  $1$ Gpc in case of the following assumptions: matter-antimatter 
symmetric Universe, continuous close contact between domains of matter and
antimatter and adiabatic perturbations~\cite{cohen98}.
This constraint  is not  applicable to
isocurvature baryogenesis models~\cite{dolgov01}, like the one discussed below, 
according to which there was not a close contact between matter and 
antimatter
regions, and afterwards  the separation  increased.
Besides, the assumption for the asymmetry is not 
obligatory!
Antimatter regions may be less than the matter ones, then gamma
observations  constrain  the antimatter-matter ratio at
different scales. 

The analysis of annihilation features within concrete baryogenesis 
 model~\cite{Khlopov}  
and the 
EGRET gamma-ray background data showed that even a small fraction
($<10^{-6}$) of antimatter stars in  our Galaxy is allowed! 
The allowed mass range $10^4-10^5M_{Sun}$ corresponds to 
  antistar globular cluster~\cite{sakharov02,belotsky00}.  
An interesting possibility of primordial antiblackholes, 
antiquasars and antistars was revealed also in ref.\cite{silk}.

And as we will discuss below, within the framework of the presented here
baryogenesis model, antigalaxies and anticlusters may be a possibility, 
too.

{\it CR and $\gamma$-ray data do
not rule out antimatter domains in the Universe.} 

{\bf Other observational signatures of  antimatter}
are  the distortion of the energy spectrum of the 
Cosmic Microwave Background Radiation and  spatial variations of 
the primordial light elements
 abundances. 
The isotropy of CMB rules out large voids between matter and antimatter 
regions during earlier time. Successful BBN  restricts the amount of
annihilation at early 
epoch and, hence, puts stringent limits on 
the fraction of antimatter 
~\cite{kurki,zeldovich}. 

So it is interesting to explore how large regions of antimatter
may be produced in the Universe and what are the observational 
signatures and constraints for them.

\section{The baryogenesis model}

There exist different inhomogeneous  baryogenesis models, which predict 
matter and antimatter regions~\cite{dolgov}.
We  discuss here the SUSY-baryogenesis
model, predicting vast  regions of antimatter, safely separated from the  
matter ones, so that the CR and gamma-ray constraints are satisfied.
It is discussed in detail in ~\cite{kc00,kc96}.
 It arises
naturally in the
{\it low temperature baryogenesis scenarios with baryon charge
condensate}~\cite{dk91,kc96}.

{\it Attractive features from the view point of
antimatter cosmology  are}:
It
does not suffer from the basic problems of antimatter cosmology 
models,
i.e. the causality problem, the annihilation catastrophe problem,
the domain walls problem, discussed in detail in ref.~\cite{steigman76}.
It can provide a natural separation mechanism of considerable  
quantities
of matter from such ones of antimatter.
The characteristic scale of antimatter regions and their distance 
from
matter ones may be
in accordance with the observational constraints for natural choice 
of
parameters.
So, the model proposes the possibility that only our vicinity is
baryonic,
 while globally the Universe may contain considerable quantities of
antibaryons and in the extreme case may be symmetric.

 The essential ingredient of the model is a baryon charged 
complex scalar field $\phi$,  present  together 
with the inflaton.  A condensate  with a nonzero baryon charge is 
formed during the
inflationary period  due to 
 enhancement of quantum fluctuations of  $\phi$ 
~\cite{inflacia}:

\begin{center}
$<\phi^2>=H^3t/4\pi^2$. 
\end{center}
$\phi$
satisfies the equation
\begin{equation}
\ddot{\phi}-a^{-2}\partial^2_i\phi+3H\dot{\phi}+
{1 \over 4}\Gamma\dot{\phi}+U'_{\phi}=0,
\end{equation}
 where $H=\dot{a}/a$.
The baryon charge of the field is not
conserved at large values of $\phi$ due to 
 B-violating self-interaction terms in the field's
potential:

\begin{equation}
U(\phi)= m^2|\phi|^2+{\lambda_1\over 2}|\phi|^4
+{\lambda_2\over 4}(\phi^4+\phi^{*4})+{\lambda_3\over 4}|\phi|^2
(\phi^2+\phi^{*2})
\end{equation}

 We have studied the evolution of the condensate after inflation
for the
 case when at the end of inflation the Universe is dominated
by a coherent oscillations of the inflaton field:

$\psi=m_{PL}(3\pi)^{-1/2}\sin(m_{\psi}t)$,
$H=2/(3t)$,  $m \ll H_I$,  $\lambda_i\sim \alpha_{GUT}$,
$\phi^{max}_o \sim H_I\lambda^{-1/4}$ and $\dot{\phi_o}=0$.

 After inflation $\phi$  oscillates around its
equilibrium
point with a decreasing amplitude, as a result of the
Universe expansion and to the particle production by the oscillating
scalar field~\cite{dk90}, due 
to the coupling to fermions $g\phi \bar{f}_1 f_2$, where $g^2/4\pi = 
\alpha_{SUSY}$. 
The amplitude of $\phi$ is damped as $\phi \rightarrow
\phi \exp(-\Gamma t/4)$ and the baryon charge, contained in the
$\phi$ condensate, is exponentially reduced also.
If $\Gamma=const$  the baryon asymmetry is waved away till
baryogenesis epoch unless  the scalar field is the inflaton
itself. 
 However,  this case is forbidden by the CMB anisotropy data. 
If  $\Gamma$ is a decreasing function of time, the baryon 
charge
contained in $\phi$ may survive until 
$B$-conservation epoch $t_b$~\cite{dk91}.
 At  $t_b$ the baryon charge is transfered to that of the quarks during
the decay of the field $\phi \to q\bar{q} l \gamma$ and  
  an antisymmetric plasma appears.
Its charge, eventually diluted further 
by some entropy
generating processes, dictates the observed baryon asymmetry.

\section{ Evolution of the baryon density
distribution}

The necessary conditions for generation of vast separated regions of 
matter and 
antimatter in the model are: initial space distribution  $\phi(r,t_0)$,
unharmonic potential and inflationary expansion. 
We studied the evolution of the baryonic space distribution, assuming  
  a monotonic initial 
distribution of the baryon density within a domain with a certain
sign of the baryon number $\phi(r,t_0)$.
For different sets of parameter values of the model
$\lambda_i$, $\alpha$,
$m/H_i$, we have
numerically followed the evolution $B(t,r)$ for
all initial values of the field $\phi_o^i=\phi(r_i,t_0)$ till 
$t_B$.  

In case of nonharmonic field's potential, the initially
monotonic space behavior is quickly replaced by space oscillations of
$\phi$, because of the dependence of the period on the amplitude~\cite{cd92}.
In our model the dependence is $\omega\sim\lambda^{1/2}\phi_i(r)$.
As a result in different
points different periods are observed and spatial behavior of $\phi$
becomes quasiperiodic. Correspondingly, the spatial distribution of baryons
  $B(t_B,r)$ at the moment of baryogenesis is found to be
quasiperiodic (Fig.~2).

\begin{figure}
\hbox{\vspace{-0.1cm}}
\mbox{\hspace{2cm}}\epsfig{figure=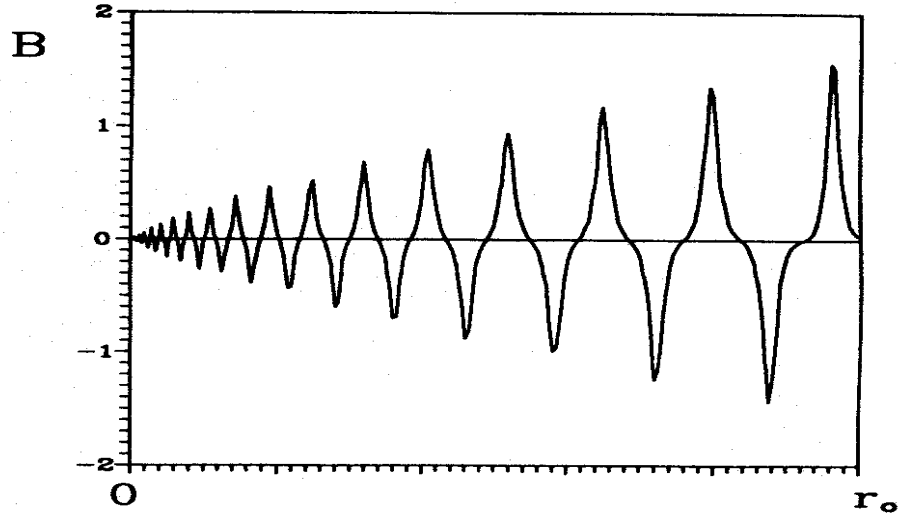,height=7cm,width=12cm}
\caption{The space distribution of baryon charge at $t_B$
 for $\lambda_1=5 \times 10^{-2}$, $\lambda_2=\lambda_3=
\alpha=10^{-3}$, $H_I/m=10^7$.
}
\hbox{\vspace{-1cm}}
\end{figure}

The region $r_0$ which initially was characterized with its baryon excess
splits into regions with baryon excess and such of baryon
underdensities~\cite{kc96}. Due to the smoothly decreasing baryon density 
towards the
borders between the  baryonic and antibaryonic regions,
predicted by the model,   annihilation is not considerable at $t_B$.
After that, the baryon and antibaryon regions further contract
towards their centers, where density is higher. Hence,
matter and antimatter domains become separated by large  empty
from baryons voids, perhaps filled with dark matter.
Thus the stringent limit~\cite{cohen98} on antimatter domains
is evaded.

Two  cases are possible:

A. {\it Stochastic CP-violation}: The variations appear around zero
baryon
charge.  The initially
baryonic domain is broken to baryonic and antibaryonic regions and
divided by nearly baryonically empty space. The case is
attractive as far as it allows the realization of symmetric
Universe without domain walls. However,  the
resulting fluctuations of the baryon density may be considerable and
 lead to unacceptably large angular variations of the
microwave background radiation.

B. {\it Stochastic$+$explicit CP-violation}:
The
field's equilibrium value is non zero, and the fluctuations
of the field around  it  result into fluctuations of the
baryon density around some mean number. Then at $t_B$ the domain
with a given sign of explicit CPV may consist predominantly
of
baryonic regions plus small quantity (for $l\sim 100$ Mpc  it is
$\sim
10^{-4}$)  of antibaryonic ones.
Though not so aesthetic, because in that case there should be besides
the stochastic CPV discussed, another mechanism of CPV
producing the mean baryon density, this case is more promising.

Due to inflation the regions with different
baryon density (overdensity, underdensity or density of
 antibaryons) become macroscopically large $d \rightarrow d\exp(Ht)$.
{\it The characteristic scale between matter and
antimatter regions}  is a function of
the models  parameters:
the coupling constants of the potential $\lambda_i$, the initial 
amplitudes
of the field $\phi(r,t_i)$, the period of baryogenesis $t_b$
 and the characteristic
scale of the baryon space variation at the inflationary stage $r_o$.
 The provided analysis showed that
 for a natural choice of the values of these parameters
the separation scale may be in the Mpc - 100 Mpc range. 

\section{Predicted antimatter structures and
observational constraints}

 Using the  constraints from gamma rays and CR data,  BBN 
and  CMB anisotropy measurements, we discuss 
different realizations of the model.

Recent CMB measurements ruled out  pure isocurvature perturbations models,
so, accordingly, the case when the baryon charge carrying field is the
inflaton itself, is excluded.  Other possibilities, when besides the
inflaton there exists a second scalar field during inflation with the
features discussed in our model remain viable~\cite{enqvist}. 
According to the recent mixed  isocurvature plus adiabatic models, although 
the isocurvature contribution is not suggested it has neither been ruled
out.

A. {\it Stochastic CP-violation}

A.1 The first most simple case we considered in~\cite{kc00}  
aiming to explain the observed $120$ Mpc periodicity in the visible matter
distribution, 
assumed that the overdensity regions correspond to 
galaxy or antigalaxy superclusters with big voids between with a 
characteristic 
size $\sim 120$ Mpc.
In that case the antimatter domains are roughly of the same scale and the
similar density as the matter ones.  CR and gamma-ray data  
constraints are fulfilled.
Large variation  of the  primordially produced elements, should
be observable at the corresponding scales. There are no data for the
rest light elements at large distances, however the observed  D
towards high-$z$  quasars shows some deviations from the expected
primordial plateau.
Alas, in that case the magnitude of the isocurvature perturbations is high 
and may induce  CMB anisotropies not compatible with the data~\cite{enqvist}.
\footnote{Calculation of the resulting angular variations of the 
temperature
of CMB  in the specific case are not done.}

A.2 Smaller  structures of 
antimatter  $< 10-20 $ Mpc are  possible. 
The CMB constraint weakens when decreasing the
scale.
However, CR and gamma-ray data restricts the number of such smaller
antimatter objects, not excluding, however, the possibility for their
existence. 
Spatial variations of the light elements are expected also.

B. {\it Stochastic$+$explicit CP-violation}:

B.1. There exist vast matter superclusters with typical scale $D$ at a 
$L\sim 120$ Mpc separation (as observed),
while the antimatter objects are of characteristic scales
$d\le 10^{-4}D$. Hence, depending on the following evolution
these antimatter regions may collapse
to form small antigalaxies, antistar clusters or vast dense antihydrogen 
clouds.
They are at a  safe distance from the matter superclusters at about 
$l_b \sim 60$ Mpc. All the observational constraints may be satisfied.

B.2  The scales of the antimatter domains are of 
 galaxy cluster or galaxy  scales.
Different
possibilities for antimatter domains may be realized,
namely between galaxy clusters  an antimatter galaxy may wonder,
in the space between groups of galaxies a globular star anticluster 
may be found.  

{\bf In conclusion}: 

 A  baryogenesis model predicting vast antimatter
regions safely separated from  the matter ones  
is discussed from the
viewpoint of existing observational and theoretical constraints
on antimatter in the Universe.
The analysis shows that    
different antimatter structures are possible:
antigalactic clusters, antigalaxies situated between clusters of 
galaxies, antistar globular clusters.  

 The discussed mechanism of separation of matter from antimatter 
domains  
 could be realized in a  variety of models, depending on the type of
the baryogenesis model, on the field potential, on the type of 
the
CP-violation, on the initial space distribution of the baryon
density at the inflationary stage, etc.

It looks
probable that the results of future experiments  on long
balloon flights and spacecraft, planning to measure antiproton and 
positron
spectra at wide range of energies ($0.1-150$ GeV) (as  by PAMELA magnetic
spectrometer) and reach a sensitivity for antinuclei at $\sim 10^{-7}$
(AMS
magnetic spectrometer), will  reveal the secrets of nearby
(well.. up to 150 Mpc)  antiworlds. It is exciting that we may know
soon the answer.
Future positive indications for antimatter
may help also to choose among the existing variety of
"anti" baryogenesis models, and for the case discussed here, it may
reveal SUSY parameters, as well.

\hbox{\vspace{-1cm}}
\section*{Acknowledgments}
D.K. thanks the  organizing committee of the conference
for the hospitality and financial support for her    
participation. D.K. appreciates useful 
discussions and suggestions by A. Dolgov and A. Sakharov, and    
 acknowledges  CERN TH visiting position and 
Abdus Salam ICTP regular associateship.

\hbox{\vspace{-1cm}}


\section*{References}
\begin{thebibliography}{99}

\bibitem{saeki98} T. Saeki  {\it et al}, \Journal{\PLB}{422}{319}{1998}. 
\bibitem{ormes97} J. Ormes  {\it et al}, {\it Ap.J.} {\bf 482}, L187
(1997).
\bibitem{steuer01} M. Steuer, \Journal{\NCA}{24}{661}{2001}.
\bibitem{alpat00} B. Alpat, \Journal{\NPB}{S85}{15}{2000}.
\bibitem{orloff} P. Chardonnet, J. Orloff, P. Salati, 
\Journal{\PLB}{409}{313}{1997}.
\bibitem{BESS}  S. Orito  {\it et al}, BESS Collaboration,  
\Journal{\PRL}{84}{1078}{2000}; Y. Asaoka  {\it et al}, BESS Collaboration, 
\Journal{\PRL}{88}{051101}{2002};
 Matsunaga H.  {\it et al}, \Journal{PRL}{81}{4052}{1998}.
\bibitem{CAPRICE}  WiZard/CAPRICE Collaboration,
 {\it Ap. J.} {\bf 561}, {787} (2001).
 \bibitem{Maeno00}T. Maeno  {\it et al}., BESS Collaboration,
{\it Astropart.Phys.} {\bf 16}, 121 (2001).
\bibitem{Bieber} J. Bieber, \Journal{\PRL}{83}{674}{1999}.
\bibitem{Bergstrom} L. Bergstrom, J. Edsjo, P. Ullio, astro-ph/9902012.
\bibitem{donato} F. Donato  {\it et al}.,  {\it Ap.J.} {\bf 563}, 172 
(2001).
\bibitem{kvp02} D. Kirilova, T. Valchanov, M.Panayotova, in
preparation.
\bibitem{boezio02} M. Boezio  {\it et al}, astro-ph/0103513, 2002.
{ormes97}
\bibitem{stecker} F. Stecker, these proceedings, hep-ph/0207323.  
\bibitem{steigman76} G. Steigman, {\it Ann. Rev. Astr. Astrop.} {\bf 
14}, 
339 (1976).
\bibitem{wolfendale} A. Dudarevich, A. Wolfendale, {\it MNRAS}{\bf 268}, 
609 (1994).
\bibitem{stecker71} F. Stecker, {\it et al}, 
\Journal{\PRL}{27}{1469}{1971}; 
 \Journal{\NPB}{252}{25}{1985}.
\bibitem{cohen98} Cohen A.  {\it et al}, {\it Ap.J.} {\bf 495}, 539 
(1998).
\bibitem{dolgov01} A. Dolgov, \Journal{\NPB}{S95}{42}{2001}, 
hep-ph/0012107.  
\bibitem{Khlopov} M. Khlopov, S. Rubin, A. Sakharov,  
\Journal{\PRD}{62}{083505}{2000}.
\bibitem{sakharov02} A. Sakharov, M. Khlopov, S. Rubin
in  {\em Budapest 2001, High energy physics}, hep2001/212, Budapest, 2001, 
astro-ph/0111524. 
\bibitem{belotsky00}K. Belotsky {\it et al}, {\it Yad.Fiz.} 
{\bf 63}, 290 (2000).
\bibitem{silk} A. Dolgov, J. Silk, \Journal{\PRD}{47}{4244}{1993}.
\bibitem{kurki} J. Rehm and K. Jedamzik, \Journal{\PRL}{81}{3307}{1998};
H. Kurki-Suonio and E. Sihvola, \Journal{\PRL}{84}{3756}{2000}.
\bibitem{zeldovich}V. Chechetkin {\it et al}, 
\Journal{\PLB}{118}{329}{1982}. 
\bibitem{dolgov} A. Dolgov, these proceedings; Z. Berezhiani,  these 
proceedings; A. Dolgov, {\it Phys.Rep.}{\bf 222}, 309 (1992).
\bibitem{dk91} A. Dolgov and D. Kirilova, {\it J.M.Phys. 
Soc.} {\bf 1}, 217 (1991).
\bibitem{kc00} D. Kirilova and M. Chizhov, {\it MNRAS} {\bf 314}, 256 
(2000). 
\bibitem{kc96}D. Kirilova and M. Chizhov, {\it Astr. Astrop. Tr.} {\bf 10}, 
69 (1996).
\bibitem{cd92} M. Chizhov and  A. Dolgov, 
\Journal{\NPB}{372}{521}{1992}.
\bibitem{inflacia}  Bunch T. and Davies P., {\it Proc. Roy.Soc.} {\bf 
A360},  117 (1978);
 A. Vilenkin and L. Ford,  \Journal{\PRD}{26}{1231}{1982};
 A. Linde 1982, \Journal{\PLB}{116}{335}{1982}.
\bibitem{dk90}A. Dolgov and D.Kirilova, {\it Yad. Fiz.} {\bf 51}, 273 
(1990).
\bibitem{enqvist}K. Enqvist, H. Kurki-Suonio, J. Valiviita,
\Journal{\PRD}{26}{103003}{2000}; M. Bucher, K. Moodley, N. Turok,
astro-ph/00212141.

\end{thebibliography}
\end{document}